\documentclass[12pt]{article}
\usepackage{times}
\usepackage{amssymb}
\usepackage{graphicx}
\begin{document}
\pagestyle{myheadings}
\markright{Christini and Kaplan}

\begin{center}
\large {\sc \bf Adaptive estimation and control of unstable periodic dynamics in excitable biological systems\\}
\bigskip
\end{center}
\begin{center}
\baselineskip=16pt
\title{}
David J. Christini$^{\dagger,}$\footnote{email: dchristi@med.cornell.edu} and Daniel T. Kaplan$^{\ddagger,}$\footnote{email: kaplan@macalester.edu}\\
\bigskip
$^\dagger$Division of Cardiology, Department of Medicine, Cornell University Medical College, New York, NY 10021\\
$^\ddagger$Department of Mathematics and Computer Science, Macalester College, 
St. Paul, MN 55105\\

\bigskip
\today
\end{center}
\bigskip

\begin{abstract}
Dynamical control of excitable biological systems is often complicated
by the difficult and unreliable task of pre-control identification of
unstable periodic orbits (UPOs). Here we show that, for both chaotic
and nonchaotic systems, UPOs can be located, and their dynamics
characterized, {\em during} control.  Tracking of system
nonstationarities emerges naturally from this approach.  Such a method
is potentially valuable for the control of excitable biological
systems, for which pre-control UPO identification is often impractical
and nonstationarities (natural or stimulation-induced) are common.
\end{abstract}

\newpage 

Chaos control techniques have been applied to a number of excitable
biological
systems~\cite{garfinkel:1992a,schiff:1994b,christini:1996a,brandt:1996a,hall:1997a}
comprised of spontaneously firing cells.  Such control typically
attempts to replace an unwanted irregular or higher-order firing
pattern with a lower-order periodic rhythm.  One particular control
technique, PPF control~\cite{garfinkel:1992a}, uses isolated
electrical stimuli to cause the cells to fire at a specified time,
thus altering the variable of interest, the inter-excitation interval.
In the idealized situation presented in Refs.~\cite{garfinkel:1992a}
and~\cite{schiff:1994b} the PPF stimuli achieve control by placing
the state of the biological system onto the stable manifold of a
desired unstable periodic orbit (UPO).

The successful application of PPF control requires an estimate of the
location of the uncontrolled system's UPO and corresponding manifolds.
UPOs and their eigenvalues~\cite{fn:eigenvalues} are typically
characterized from measurements of the system in free-running mode,
without external
stimulation~\cite{garfinkel:1992a,schiff:1994b}. PPF-type~\cite{fn:ppf-type}
stimuli are then used to alter the inter-excitation interval in an
attempt to place the system state point onto the estimated stable
manifold and therefore stabilize the UPO.

Proper estimation of the UPO and its characteristics is of fundamental
importance to PPF-type control for several reasons. First, optimal
control (which we consider to be stabilization of the desired UPO with
a minimal number of stimuli) is achieved when the system state is
placed onto the stable manifold. Without placement directly onto a
stable manifold, control can be achieved via alternative dynamical
mechanisms~\cite{christini:1995c,glass:1994a,sauer:1997a,christini:1997b,kaplan:1999a}
that require more frequent stimulation. Second, knowledge of unstable
orbits can provide a skeleton upon which to build a model of the
overall system.  Third, the UPO may change in time; by continuously
tracking the properties of the orbit, one can adaptively change the
control parameters in order to maintain the controlled stability of
the orbit.  Changes in the UPO may stem from autonomous drifts in the
properties of the biological system, or may be a response to the
control stimuli (a tissue's dynamical or electrochemical properties
may be modified by stimulation~\cite{kunysz:1995b,wanzhen:1991a}).
With these reasons in mind, this report is concerned with ways to use
PPF-type control in order to identify, control, and track such UPOs
and their manifolds.

The detection of a UPO using data collected in the uncontrolled,
free-running system~\cite{garfinkel:1992a,schiff:1994b} can be
problematic.  As mentioned above, one problem with such an approach is
that excitable biological systems are commonly nonstationary. Thus, a
pre-control UPO estimate may become invalid before or during the
control stage. Another problem is that in the free-running system, the
system's state may spend little time in the vicinity of the orbit.  As
an extreme case, a system with a stable attracting periodic orbit may
well have other UPOs, but the free-running system will visit only the
stable orbit.  Even with a lengthy free-running data collection stage,
there is no guarantee that there will be sufficient data from within the
UPO neighborhood; a paucity of such data from within the UPO
neighborhood renders dubious the reliability of the estimated UPO
characteristics.  Statistical tests have been proposed to validate UPO
existence~\cite{so:1997a}, but the most compelling evidence comes when
control of a putative orbit is successfully achieved. With this in
mind, an alternative to pre-control UPO identification is to locate
and characterize a UPO while attempting control. 

To this end, Kaplan~\cite{kaplan:1999a} recently showed that control
can be used to locate UPOs by tuning the control parameters to be near
a bifurcation in the controlled system's dynamics.  Such control
allows the experimentalist to circumvent the major hurdle of
pre-control UPO identification: the limited time that the state point
spends in the UPO neighborhood. However, although the system's state
stays near the UPO during such control, estimation of UPO
characteristics is complicated by the fact that the natural UPO
dynamics are masked by the control stimuli.  So, the experimentalist
faces a choice of studying the free-running system with infrequent UPO
data, or studying the controlled system with plentiful data but with
obscured or altered dynamics.  As we show in this report, the latter
alternative can be rendered feasible by circumventing the masking in
either of two ways: 1) estimating the natural UPO dynamics {\em
between} intermittently applied stimuli or 2) jiggling the control
parameters.

We will examine systems whose natural, uncontrolled dynamics can be
approximated by an autoregressive system $x_{n+1} = f( x_n, x_{n-1})$.
Such systems can have many types of UPOs.  For control of excitable
biological systems, $x_n$ is taken to be the time interval between the
$n^{th}$ firing of the system and the previous firing.  As long as the
state point $(x_n, x_{n-1})$ is in the neighborhood of a UPO, the
system dynamics can be approximated linearly as $x_{n+1} = a x_n +
bx_{n-1} + c$, or, rewriting the constants $a$, $b$, and $c$ in terms
of the parameters of the UPO:
\begin{equation}
\label{eq:dyn-linear}
x_{n+1} = 
(\lambda_s + \lambda_u) x_n - \lambda_s \lambda_u x_{n-1} +
x_{\star} ( 1 + \lambda_s \lambda_u - \lambda_s - \lambda_u)
\end{equation}
where $\lambda_s$ and $\lambda_u$ are the eigenvalues of the
linearized system, and $x_{\star}$ is the location of the UPO sampled
once per cycle.  The notation is intended to suggest that there is one
stable and one unstable eigenvalue, as required for a saddle-type
fixed point.  However, the equation is applicable even when both
eigenvalues are unstable.

The application of PPF-type control changes the dynamics near the UPO to
a nonlinear form:
\begin{equation}
\label{eq:ppf-basic}
x_{n+1} = \mbox{min} \left\{
\begin{array}{ll}
(\lambda_s + \lambda_u) x_n - \lambda_s \lambda_u x_{n-1} +
x_{\star} ( 1 + \lambda_s \lambda_u - \lambda_s - \lambda_u) &\mbox{natural dynamics} \\
\widehat{\lambda}_s ( x_{n} - \widehat{x}_\star ) + \widehat{x}_\star & \mbox{control stimulus} 
\end{array}
\right.
\end{equation}
where $\widehat{x}_\star$ and $\widehat{\lambda}_s$ are estimates of
the UPO position $x_\star$ and stable manifold slope $\lambda_s$.
Kaplan~\cite{kaplan:1999a} showed that for a flip-saddle, when
$\widehat{x}_\star \approx x_\star$ and $|\widehat{\lambda}_s|<1$ the
controlled system modeled by Eq.~\ref{eq:ppf-basic} will be
characterized by a control stimulus applied either every interval
(when $\widehat{x}_\star < x_\star$) or every second interval with the
intervening intervals being terminated naturally (when
$\widehat{x}_\star > x_\star$).  As shown in Ref.~\cite{kaplan:1999a},
$x_\star$ can be located by systematically scanning over a range of
$\widehat{x}_\star$ for the stimulus-pattern bifurcation.

In the system of Eq.~\ref{eq:ppf-basic}, because the natural dynamics
are obscured by the control stimuli, it is not possible to estimate
$\lambda_s$ and $\lambda_u$.  For the case where $\widehat{x}_\star <
x_\star$, (when control stimuli are applied every interval), the
controlled dynamics are simply
\begin{equation}
\label{eq:pacing}
 x_{n+1} = \widehat{\lambda}_s (x_{n} - \widehat{x}_\star ) + \widehat{x}_\star.
\end{equation}
The natural $\lambda_s$ and $\lambda_u$ do not enter into these
dynamics and therefore cannot be estimated from them.  For the case
where $\widehat{x}_\star > x_\star$ (when control stimuli are applied
every second interval), the controlled dynamics for intervals that end
naturally without a control stimulus, are
\begin{equation}
\label{eq:lumped}
x_{n+1} = \bigl(\widehat{\lambda}_s ( \lambda_s + \lambda_u ) - \lambda_s \lambda_u\bigr)
x_{n-1} + 
\bigl(1 + \lambda_s \lambda_u - \widehat{\lambda_s}(\lambda_s+\lambda_u)\bigr) \widehat{x_\star}
\end{equation}
(as can be found by substituting the bottom equation of
Eq.~\ref{eq:ppf-basic} for $x_n$ in the top equation).  Note that from
measurements of $x_{n+1}$ and $x_{n-1}$, only the lumped constant
parameter 
$\lambda_s \lambda_u - \widehat{\lambda_s}(\lambda_s + \lambda_u)$ 
in Eq.~\ref{eq:lumped} can be estimated, and not
$\lambda_s$ and $\lambda_u$ individually.

Christini and Collins~\cite{christini:1997b} proposed a simplified
modification of PPF control, which they termed SMP control.  They
showed that effective control can be accomplished by turning off the
control stimuli and allowing the system to free-run according to the
natural dynamics until the state point $(x_{n}, x_{n-1})$ wanders out
of the UPO neighborhood.  Only when the difference between $(x_{n},
x_{n-1})$ and the UPO $(x_\star, x_\star)$ reaches a threshold is
control reactivated to return the state point to the UPO neighborhood
via the stable manifold.  In Ref.~\cite{christini:1997b}, the primary
motivation for such intermittent perturbation was to minimize control
interventions in order to limit stimulation-induced modification of
the dynamical or electrochemical properties of the excitable
tissue~\cite{kunysz:1995b,wanzhen:1991a}.  In the present context,
intermittent stimulation provides another important benefit: allowing
observation and characterization of the natural UPO dynamics between
control perturbations.

When the control stimuli are turned off when using the SMP strategy,
the system dynamics near the fixed point are given by
Eq.~\ref{eq:dyn-linear}.  The parameters $\lambda_s$, $\lambda_u$, and
$x_\star$ can then be estimated by linear regression of $x_{n+1}$ on
$x_n$ and $x_{n-1}$.  We define a ``natural triplet'' to be 
a sequence $(x_{n+1}, x_n,
x_{n-1} )$ in which interval $n+1$ is terminated naturally, but
intervals $n$ and $n-1$ could be terminated naturally or by control
stimuli. Only natural triplets can be used in the regression.

In order to track UPO drift, estimates of $\lambda_s$, $\lambda_u$,
and $x_\star$ are made from the last $N$ natural triplets $(x_{n+1},
x_n, x_{n-1} )$.  In this letter, we take $N=10$.  After each
estimation, the control parameters $\widehat{\lambda}_s$ and
$\widehat{x}_\star$ in Eq.~\ref{eq:ppf-basic} are adjusted
accordingly.

Care must be taken when performing the linear regression.  If only one
eigenvector is required to characterize the data fit to
Eq.~\ref{eq:dyn-linear}, then the parameter estimations will not
accurately represent the natural two-manifold UPO dynamics. This
situation occurs when one of the manifolds has little or no influence
on the state dynamics for several consecutive natural triplets. For
example, when control is turned off when using the SMP strategy, the
state point will tend to retreat from the UPO along the unstable
manifold.  Thus, the natural dynamics will be $x_{n+1} = \lambda_u (
x_{n} - x_\star) + x_\star$, which does not reflect $\lambda_s$.  If
the $N$ points used in the estimation consist mainly of such points,
the parameter estimations will not accurately represent the natural
two-manifold UPO dynamics.  Such a situation can be detected by using
singular value decomposition (SVD)~\cite{press:1992a} to carry out the
linear regression: if the ratio between the regression's largest and
smallest singular values is exceedingly large, the estimate is
dubious.  If this is the case, $\widehat{x}_\star$ and
$\widehat{\lambda}_s$ from the last valid estimation of
Eq.~\ref{eq:dyn-linear} should be used for setting the control
parameters in Eq.~\ref{eq:ppf-basic}.

We illustrate the SMP characterization and tracking technique using
the chaotic H\'enon map,
\begin{equation}  
x_{n+1}=1.0-Ax_{n}^2+Bx_{n-1},
\label{eq:henon}
\end{equation}
where $A=1.4$, $B=0.3$, and $x_n$ represents the $n^{{th}}$
inter-excitation interval. With these parameter values, the system is
chaotic and has a flip-saddle UPO at $x_\star=0.8839$, with
$\lambda_u=-1.9237$ and $\lambda_s=0.1559$.
Figure~\ref{fig:rtsmp_hen} shows a trial demonstrating the adaptive
estimation and control of this UPO.  Initially, the H\'enon map was
free-run for 100 points without control
[Fig.~\ref{fig:rtsmp_hen}(a)]. At $n=100$, control was activated,
setting the control parameter $\widehat{\lambda}_s = 0$ and scanning
for ${x}_\star$ by systematically increasing $\widehat{x}_\star$.  For
$\widehat{x}_\star < x_\star$ the resulting controlled dynamics show a
fixed point at $\widehat{x}_\star$ with the control stimulus being
applied at every interval. At $\widehat{x}_\star = x_\star$, a
period-doubling bifurcation occurs, thus marking the location of the
flip-saddle UPO.  As shown in the inset of
Fig.~\ref{fig:rtsmp_hen}(b), the bifurcation occurred at $n=174$.

Following detection of the bifurcation, control continued with
$\widehat{x}_\star$ set to $0.9073$, the midpoint of the last
pre-bifurcation pacing interval and the first post-bifurcation pacing
interval. After $N$ natural triplets had occurred [see inset in
Fig.~\ref{fig:rtsmp_hen}(b)], the first SVD estimation of $\lambda_s$,
$\lambda_u$, and $x_\star$ was performed~\cite{fn:173-180}. After
$n=190$, control followed the SMP protocol, with control stimuli used
only when $|x_\star - x_n| >
\delta$.  ($\delta$ was set to $0.001$ for all trials in this
study.) SMP successfully stabilized the UPO with control stimuli being
provided approximately every 20$^{th}$ interval as seen in
Fig.~\ref{fig:rtsmp_hen}(c).  Figure~\ref{fig:rtsmp_hen}(d) shows that the
control-stage SVD estimates $\widehat \lambda_s$ and $\widehat
\lambda_u$ (re-estimated via SVD following every natural interval using the
most recent $N$ natural triplets) were close to their true values.

After $n=500$, to simulate a noisy system, a Gaussian white noise
iterate (standard deviation 0.0001) was added to each non-controlled
H\'enon map iterate.  In the noisy system, control required more
frequent SMP perturbations because the additive noise caused the
system state point to wander more quickly away from $x_\star$.  Due to
the additive noise, $\widehat \lambda_s$ and $\widehat \lambda_u$
fluctuated, but remained scattered around the true values
[Fig.~\ref{fig:rtsmp_hen}(d)].

This technique can also be used to locate and stabilize UPOs in
nonchaotic systems. We performed a trial controlling the H\'enon map
of Eq.~\ref{eq:henon} with $A=1.0$ and $B=0.3$. With these parameter
values, the system settles into a stable period-4 rhythm.  However,
there is an underlying unstable flip-saddle UPO at $x_{\star}=0.7095$,
with $\lambda_u=-1.6058$ and $\lambda_s=0.1868$.  This UPO cannot be
detected from free-running data, but the bifurcation search, SVD
parameter estimation, and SMP control were able to locate and
stabilize the UPO. For this trial, the bifurcation search, control
perturbations, and manifold estimations were all qualitatively similar
to those shown for the chaotic H\'enon map in
Fig.~\ref{fig:rtsmp_hen}. As in Fig.~\ref{fig:rtsmp_hen}, control
remained effective when a Gaussian white noise iterate (standard
deviation 0.0001) was added to each non-controlled H\'enon map
iterate. Given the prevalence of pathologic nonchaotic rhythms in
excitable biological
systems~\cite{christini:1996a,brandt:1996a,hall:1997a}, this example
demonstrates an important capability of this control technique.

It is of particular interest to be able to track drifting UPOs in
nonstationary systems.  To illustrate how this can be done, we use the
H\'enon map with a randomly drifting parameter:
\begin{equation}  
x_n=1.0-(A+\eta_n)x_{n-1}^2+Bx_{n-2},
\label{eq:henon_ns}
\end{equation}
where $\eta_n$ is an iterate of a correlated noise
process~\cite{mannella:1989}, given by $\eta_n = 0.999\eta_{n-1} + 4.5
\times 10^{-7}
\zeta_n$, where $\zeta_n$ is Gaussian white noise with unity standard
deviation.  Figure~\ref{fig:rtsmp_ns}(a) shows the inter-excitation
intervals for a control trial of this nonstationary
system. Figure~\ref{fig:rtsmp_ns}(b), (c), and (d) show the
analytically-determined $x_{{\star}_n}$, $\lambda_{s_n}$, and
$\lambda_{u_n}$, respectively, and their SVD estimates $\widehat
x_{{\star}_n}$, $\widehat \lambda_{s_n}$, and $\widehat
\lambda_{u_n}$, respectively. These panels
demonstrate that the repeated SVD estimation was able to effectively
track the drifting parameters~\cite{fn:two_branches}.

A second method for estimating $\lambda_s$ and $\lambda_u$ is
applicable when the natural unstable dynamics are sufficiently strong
that SMP control cannot be practically applied.  As shown in
\cite{kaplan:1999a}, it is not necessary for the control parameters
$\widehat{\lambda}_s$ and $\widehat{x}_\star$ to match the natural parameters
$\lambda_x$ and $x_\star$ in order to accomplish successful control.
In the case of a flip saddle, for example, by setting the control
parameter $\widehat{x}_\star$ slightly larger than the true fixed
point location $x_\star$, the controlled system will have a period-2
orbit, where control stimuli are provided every second interval.  By
jiggling the control parameters in a small range around their nominal
values, one eliminates the linear degeneracy of Eq.~\ref{eq:pacing}
and enables $\lambda_s$ and $\lambda_u$ to be separately estimated
from the natural triplets $(x_{n+1}, x_n, x_{n-1})$.

The techniques presented in this study dispense with pre-control data
analysis and enable control of nonstationary UPOs in chaotic and
nonchaotic systems. Thus, they are more appropriate than previous
techniques for control of excitable biological systems. This fact,
coupled with experimental evidence that model-independent control
techniques can modify or eliminate pathological excitation
patterns~\cite{garfinkel:1992a,hall:1997a}, implies that they may have
clinical utility. As one possibility, we note that in some clinical
applications of tachycardia pacing, one uses rapid pacing in
order to capture the tissue's rhythm, and then gradually slows pacing
to return the heart to an acceptably slow rhythm.  The techniques
described here may be useful in maintaining capture of the rhythm
while the pacing rate is slowed.

While this study demonstrates the feasibility of controlling {\em
models} of excitable biological systems, important questions regarding
the physiological feasibility of control of {\em real} excitable
biological systems remain unanswered. One question is whether or not
excitable biological systems actually contain UPOs. This is a topic of
considerable research and
debate~\cite{pierson:1995a,so:1997a,so:1998a,petracchi:1997a}. Another
question is whether SMP control stimuli, which are large perturbations
to the electrochemical properties of the system, actually modify the
underlying UPO dynamics
~\cite{christini:1997b,kunysz:1995b,wanzhen:1991a}.  Further
investigation is needed to address these, and other, issues to
determine whether such control is physiologically feasible and
clinically useful.

\section*{ACKNOWLEDGMENTS}
\noindent
This work was supported in part by a grant from the National
Institutes of Health (R01 HL56139) [DJC] and sabbatical support
from Macalester College [DTK].

\newpage


\begin{thebibliography}{10}
\bibitem{fn:173-180}
Note that in contrast to the explanation accompanying Eq.~\ref{eq:lumped},
  estimation was possible during alternation between natural- and
  stimuli-terminated intervals due to the transient from $n=173$ to $n=180$.
\end{thebibliography}


\newpage

\begin{figure}[!ht]
	\caption{A trial controlling the chaotic H\'enon map of
	Eq.~\ref{eq:henon}.  (a), (b), the inset in (b), and (c) show
	the intervals $x_n$ versus interval number $n$ for various
	stages of the control trial. Natural intervals are shown as
	filled circles, while control-induced intervals are shown as
	open triangles. (d) shows the SVD estimates (re-estimated
	following every non-control interval) $\widehat \lambda_s$ and
	$\widehat \lambda_u$ during the control stage.}
\label{fig:rtsmp_hen}
\end{figure}

\begin{figure}[!ht]
	\caption{A trial controlling the modified H\'enon map of
	Eq.~\ref{eq:henon_ns}.  (a) and the inset in (a) show the
	intervals $x_n$ versus interval number $n$ for the entire
	trial. Natural intervals are shown as filled circles, while
	control-induced intervals are shown as open triangles. (b),
	(c), (d), and the inset in (d) show the
	analytically-determined (open circles) $x_{{\star}_n}$,
	$\lambda_{s_n}$, and $\lambda_{u_n}$, respectively, and their
	SVD estimates (closed circles behind the open circles)
	$\widehat x_{{\star}_n}$, $\widehat \lambda_{s_n}$, and
	$\widehat \lambda_{u_n}$, respectively.}
\label{fig:rtsmp_ns}
\end{figure}


\clearpage

\newcounter{fig_counter}
\setcounter{fig_counter}{0}

\newpage
\enlargethispage*{1000pt}
\centerline{\includegraphics[height=8.75in,viewport=70 10 560 775,clip]{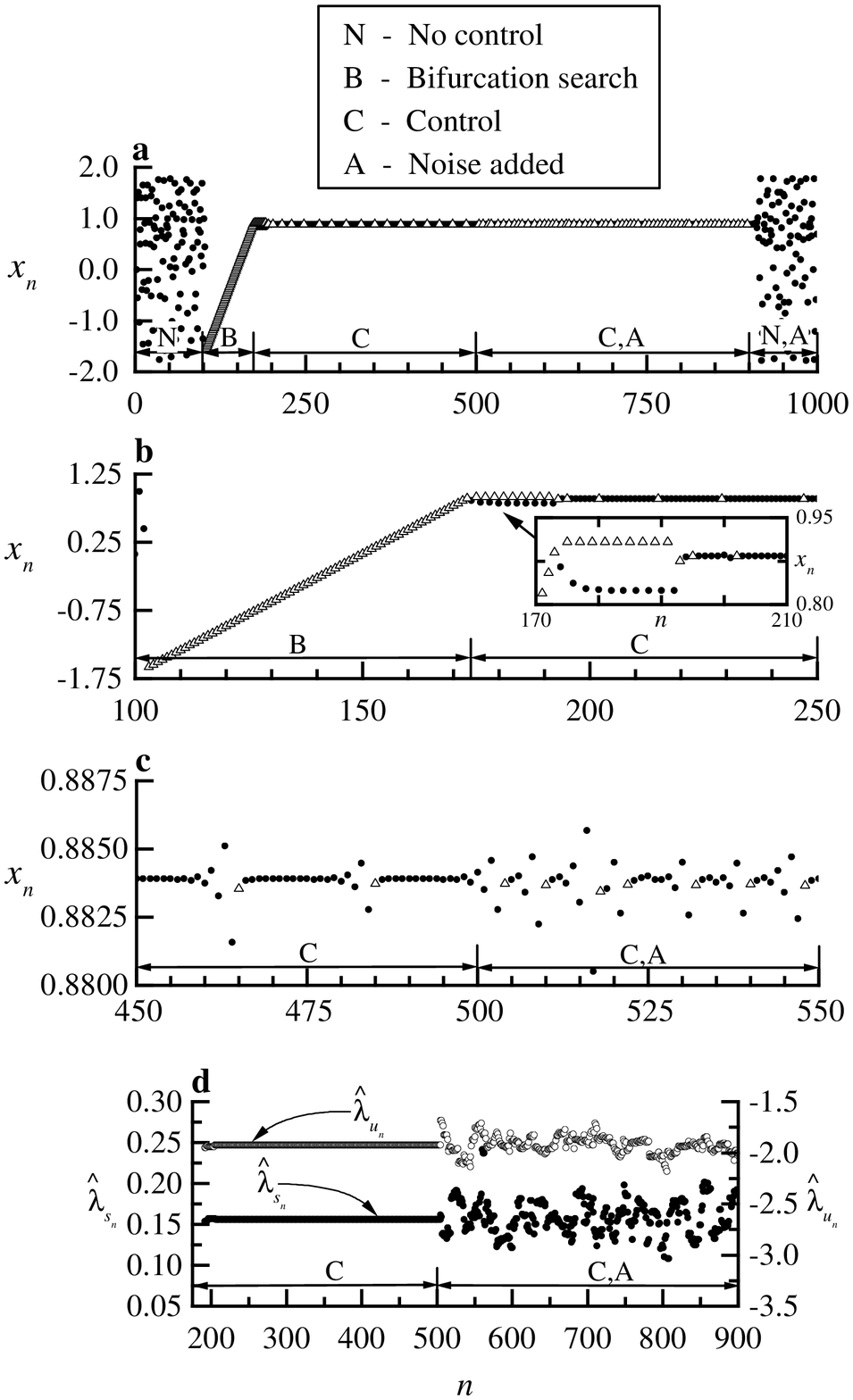}}
\stepcounter{fig_counter}
\begin{center}
\Large Fig. \arabic{fig_counter}\\
\end{center}

\newpage
\enlargethispage*{1000pt}
\centerline{\includegraphics[height=8.75in,viewport=70 10 560 750,clip]{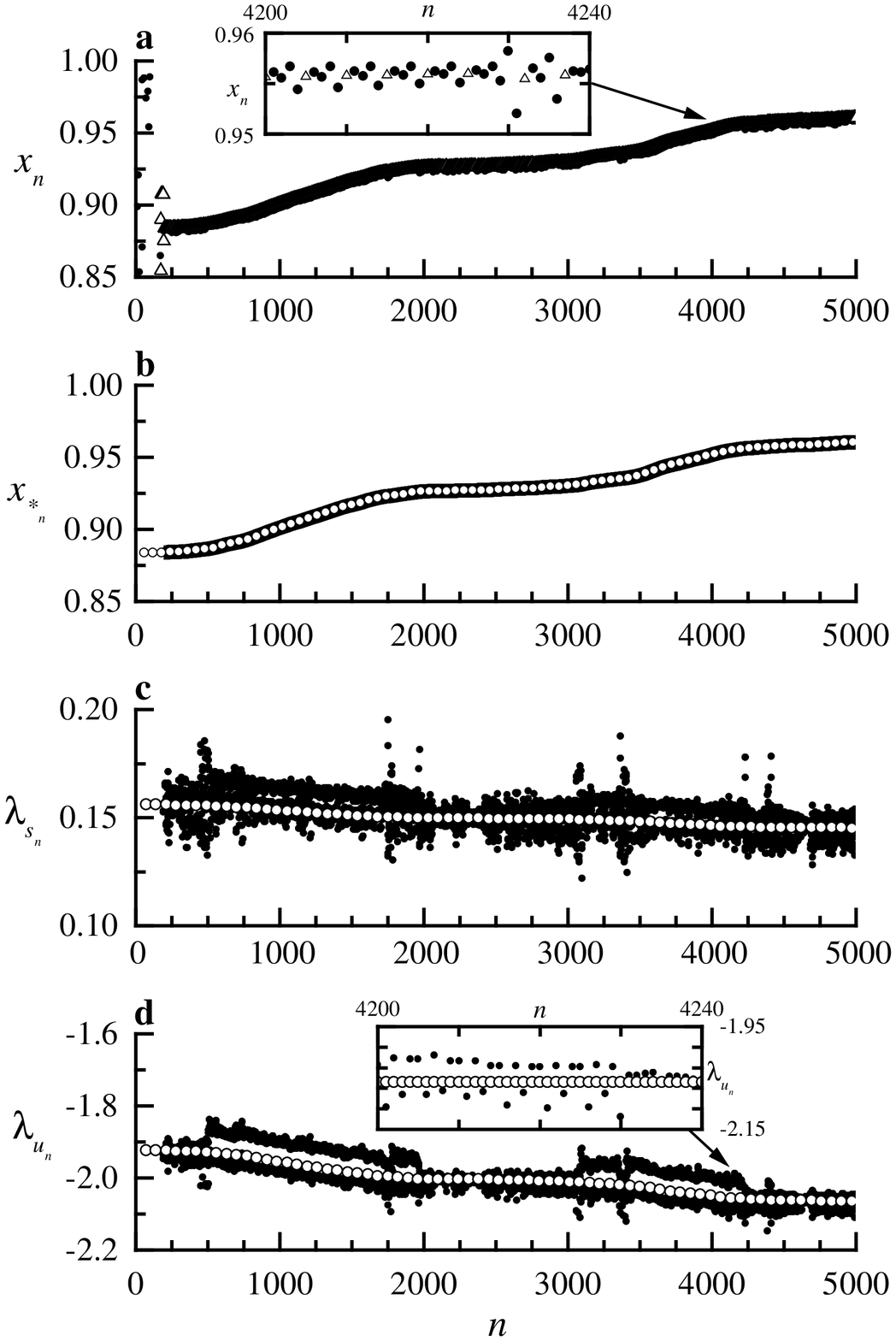}}
\stepcounter{fig_counter}
\begin{center}
\Large Fig. \arabic{fig_counter}\\
\end{center}

\end{document}